\documentclass[12pt]{iopart}

\usepackage{amssymb}
\usepackage{epsfig}
\usepackage{color}
\usepackage{adjustbox}
\usepackage{float}
\usepackage{verbatim}
\usepackage[bookmarks]{hyperref}
\usepackage{color}
\expandafter\let\csname equation*\endcsname\relax
\expandafter\let\csname endequation*\endcsname\relax
\usepackage{amsmath}
\usepackage{amsfonts}
\usepackage{enumerate}
\usepackage{amsthm}
\usepackage{graphicx}
\usepackage[english]{babel}
\usepackage{hyperref}
\usepackage{cite}
\usepackage{braket}
\usepackage[normalem]{ulem}

\begin{document}
\title[]{Entanglement in bipartite X-states: Analytical results for the volume of states with positive partial transpose}
\author{Yaqing Xy Wang}
\address{Forschungszentrum J\"ulich, Institute of Quantum Control (PGI-8), 52425 J\"ulich, Germany}
\address{Institute for Theoretical Physics, University of Cologne, 50937 Cologne, Germany}
\ead{yaq.wang@fz-juelich.de}
\author{J\'ozsef Zsolt Bern\'ad\footnote[7]{Author to whom any correspondence should be addressed.}}
\address{Forschungszentrum J\"ulich, Institute of Quantum Control (PGI-8), 52425 J\"ulich, Germany}
\ead{j.bernad@fz-juelich.de}

\date{today}

\begin{abstract}
      We provide an analytical formula for the volume  ratio between bipartite X-states with positive partial transpose and all bipartite X-states. The result applies to arbitrary $m \times n$-bipartite systems and the volume expressions are derived with respect to the Hilbert-Schmidt measure.     
\end{abstract}
\vspace{2pc}

\maketitle

\section{Introduction}

Entanglement is one of the most important features of quantum mechanics \cite{Peres, Bengtsson}. Therefore, the question of how many entangled states are in the set of all quantum states holds great significance from both philosophical and experimental points of view. The first challenge to solve this question is the search for necessary and sufficient criteria, which are capable of identifying, whether a state is separable or entangled. The necessary and sufficient criteria based on entanglement witnesses, positive maps \cite{HorodeckiRev}, and the projective cross norm \cite{Rudolph, Arveson} are computationally unusable. The other approaches are formulated in the form of simple practical algebraic tests \cite{Bengtsson, HorodeckiRev, Simnacher}, like the well-known Peres-Horodecki criterion \cite{Peresp, Horodecki} based on the partial transposition of density matrices. Following the results of St\o{}rmer\cite{Stormer} and Woronowicz \cite{Woronowicz}, this criterion completely characterizes the set of separable quantum states only for $2 \times 2$ (qubit-qubit) and $2 \times 3$ (qubit-qutrit) systems. In larger bipartite systems, there exist entangled states with positive partial transpose (PPT), and they are called bound entangled states or entangled PPT states \cite{Horodecki2}. The study of the typicality of bipartite PPT states was initiated by Ref. \cite{Sanpera}, where the volume ratio between PPT quantum states and all quantum states was numerically investigated. They have shown that this volume ratio decreases exponentially and tends to zero with increasing dimension of the quantum systems\cite{infinite}, unfortunately, there are some numerical mistakes in the paper, which have been recognized later in the literature \cite{Slater1, Slater2, Shang}. Recently, one of us has shown that the exponential character of the volume ratio is correct \cite{Sauer}, but it is faster than how was predicted by Ref. \cite{Sanpera}. 

So far extensive numerical results have been obtained on the typicality of PPT quantum states, but the analytical formulas are rare. A well-studied case is the qubit-qubit system, where all PPT quantum states are separable. Now, we summarize the analytical results obtained for several families of two-qubit states endowed with the Hilbert-Schmidt measure. In the case of Werner states \cite{Werner}, the volume ratio is $1/2$, while this value is also $1/2$ for Bell-diagonal states, where the separable quantum states form an octahedron and the whole convex set of Bell-diagonal states is a tetrahedron \cite{Horodecki3, Ziman, Leinaas}. One family of states is the so-called X-states \cite{Ravi}, because of the visual appearance of the density matrix, and can be described by seven real parameters. The volume ratio between separable X-states and all X-states is analytically known to be $2/5$ \cite{Milz}, but a similar result was obtained earlier for a parametrization of two-qubit density matrices with seven real numbers in section $2.2.3.$ of Ref. \cite{Slater1}. Another family of quantum states which has an analytical volume ratio contains real-valued two-qubit states. The ratio $29/64$ was conjectured by Ref. \cite{Slater3} and proved by Ref. \cite{Lovas}. The general two-qubit case was solved only recently. The long-standing conjecture of $8/33$ \cite{Slater2} was proved by Ref. \cite{Huong} with the help of the Duistermaat-Heckman measure, which differs from the Hilbert-Schmidt measure by a constant. Despite these efforts, there are no analytical results for bipartite quantum systems with Hilbert space $\mathbb{C}^m \otimes \mathbb{C}^n$ where $m,n > 2$. In this paper, we investigate a general bipartite quantum system for the special X-states. This is a generalization of the approach presented in Ref. \cite{Milz} and sheds some light on the role of the dimension of the quantum system characterized by $m$ and $n$. 

The paper is organized as follows. In section \ref{sec:preliminaries}, we introduce some notation, the general framework for bipartite X-states, and useful integrals. In section \ref{sec:Volumetotal}, we present the volume of the X-state space by using the  Hilbert-Schmidt measure. This is followed by our main results in section \ref{sec:VolumePPT} on the volume of those X-states, which fulfill the positive partial transpose criterion. Some remarks on the obtained result are presented in section \ref{sec:conluding}.

\section{Preliminaries}
\label{sec:preliminaries}

\subsection{Bipartite X-states}
In this subsection, we introduce bipartite X-states and derive expressions for their eigenvalues. This will prove helpful in constructing the Hilbert-Schmidt measure from the density matrix parameters and further formulate the state space volume expression. Bipartite systems consist of two quantum subsystems each consisting of several quantized levels. Mathematically, let $\mathbb{C}^m$ denote the $m$-dimensional Hilbert space of the $m$-level subsystem and $\mathbb{C}^n$ of the other $n$-level subsystem. If we fix a canonical orthonormal basis, 
then all linear operators on $\mathbb{C}^m \otimes \mathbb{C}^n$ are $mn \times mn$ matrices in $M_{mn}(\mathbb{C})$. The unique matrix $A^\dagger$ satisfying $\langle A^\dagger x, y \rangle=\langle x, Ay \rangle$ for 
all $x,y$ in $\mathbb{C}^m \otimes \mathbb{C}^n$ is called the adjoint of $A$. 
If $A=A^\dagger$, then $A$ is a self-adjoint/Hermitian matrix. We denote by $A\geq 0$, when $A$ is a positive semidefinite matrix, \textit{i.e.}, $\langle x, Ax \rangle \geq 0$ for all  $x$ in $\mathbb{C}^m \otimes \mathbb{C}^n$. 
This property implies that a positive semidefinite matrix is self-adjoint. For any matrix $A$ and any orthonormal basis $\{x_1,x_2,\dots,x_{mn}\}$ the quantity $\sum^{mn}_{i=1} \langle x_i, Ax_i \rangle$ is 
independent of the basis and called the trace of $A$, \textit{i.e.}, $\mathrm{Tr}\{A\}$. A quantum state $\rho$ on $\mathbb{C}^m \otimes \mathbb{C}^n$ is a positive semidefinite matrix with unit trace
\begin{equation}
 \rho \geq 0, \quad \mathrm{Tr}\{\rho\}=1. \nonumber
\end{equation}

In the next step, we pick an orthonormal basis, e.g., the canonical one. Bipartite X-state as $mn$ density matrices have only non-zero diagonal and complex antidiagonal entries. Thus their definition is basis dependent. Furthermore, they form a subset of all bipartite density matrices. They are the next-to-trivial class of density matrices regarding symmetry, whereas the trivial class would be diagonal matrices. In the case $m=2$ and $n=3$, an X-state reads
\begin{equation}
\label{eq:Xstateexample}
    \begin{bmatrix}
    \rho_{11} & 0 & 0 & 0 & 0 & \rho_{16}\\
    0 & \rho_{22} & 0 & 0 & \rho_{25} & 0\\
    0 & 0 & \rho_{33} & \rho_{34} & 0 & 0\\
    0 & 0 & \rho_{43} & \rho_{44} & 0 & 0\\
    0 & \rho_{52} & 0 & 0 & \rho_{55} & 0\\
    \rho_{61} & 0 & 0 & 0 & 0 & \rho_{66}\\
    \end{bmatrix}
\end{equation}
The non-zero parameters satisfy the properties mentioned above: 
\begin{equation}
 \mathrm{Tr}\{\rho\}=\sum_{i=1}^6 \rho_{ii} = 1,   
\end{equation}
and the Hermitian property, \textit{i.e.}, $\rho = \rho^{\dagger}$, implies
\begin{equation}
 \rho_{ii} \in \mathbb{R},\, \rho_{16} = \rho_{61}^*,\,\rho_{25} = \rho_{52}^*, \, \text{and} \,\, \rho_{34} = \rho_{43}^*.
\end{equation}
In the subsequent subsection, we characterize the positive semidefinite property of this family of states.

\subsection{Eigenvalues of bipartite X-states}

Given the characteristic polynomial method,
the eigenvalues of $\rho$ are values of $\lambda$ that satisfy the equation $\det(\rho - \lambda I_{mn}) = 0$, where $I_{mn}$ is the identity matrix on $\mathbb{C}^m \otimes \mathbb{C}^n$. The eigenvalues of $m \times n$ X-states can be readily calculated by noticing that the matrix $\rho - \lambda I_{mn}$ can be manipulated into an upper triangular matrix by scalar multiplication and addition of rows, where both operations preserve the determinant of a matrix. Then, the determinant becomes directly the product of all resulting diagonal elements. For example, to eliminate $\rho_{61}$ from $\rho_{2\times 3}^X - \lambda I_{6}$, where $\rho_{2\times 3}^X$ is the matrix stated in expression \eqref{eq:Xstateexample}, one can simply add $-\frac{\rho_{61}}{\rho_{11}-\lambda}$ times row one to row six, making row six equal to
\begin{equation}
\begin{bmatrix}
    0 & 0 & 0& 0&0&\rho_{66}-\lambda - \frac{\rho_{61}\rho_{16}}{\rho_{11}-
\lambda}
\end{bmatrix}    
\end{equation}
which now enters in the polynomial $\det(\rho - \lambda I_{6})$ as a factor of $\rho_{66}-\lambda - \rho_{61}/(\rho_{11}-
\lambda)$, because $\rho_{56}=0$. Now, solving for the zeros of the characteristic polynomial gives a pair of eigenvalues as solutions to 
\begin{equation}
 \lambda = \rho_{66}- \frac{\rho_{61}\rho_{16}}{\rho_{11}-
\lambda},   
\end{equation}
or
\begin{equation}
\lambda^2 - \lambda (\rho_{11} + \rho_{66}) + \rho_{66}\rho_{11} - \rho_{61}\rho_{16}= 0,    
\end{equation}
leading to the condition $\rho \geq 0$ to be translated to
\begin{equation}
(\rho_{11} + \rho_{66})^2 \geq (\rho_{11} + \rho_{66})^2 - 4(\rho_{66}\rho_{11} - \rho_{61}\rho_{16}).   
\end{equation}
Thus, we have
\begin{equation}
 \rho_{66}\rho_{11} \geq \rho_{61}\rho_{16} =\rho_{16}\rho_{16}^* = \operatorname{Re}(\rho_{16})^2 + \operatorname{Im}(\rho_{16})^2.   
\end{equation}
The general result is immediate for a $m \times n$ X-state and the expression for the eigenvalues reads
\begin{equation}
\lambda^{(i+j = mn+1)}_{\pm} = \frac{1}{2}\left[\rho_{ii}+\rho_{jj} \pm \sqrt{(\rho_{ii} + \rho_{jj})^2 - 4(\rho_{ii}\rho_{jj} - \rho_{ij}\rho_{ji})}\right].    
\end{equation}
This results in the following constraint
\begin{equation}
\label{eq:posconstraint}
\rho_{ii}\rho_{jj} \geq |\rho_{ij}|^2 = \operatorname{Re}(\rho_{ij})^2 + \operatorname{Im}(\rho_{ij})^2 \quad \text{for} \quad i+j=mn+1.    
\end{equation}

\subsection{Positive Partial Transpose} 

Let us consider the transposition map $\mathrm{\tau}_{m}$ on $M_{m}(\mathbb{C})$ and the identity operation $\mathbb{I}_{n}$ on $M_{n}(\mathbb{C})$.  Then, the partial transposition map $\rho \rightarrow (\mathrm{\tau}_{m}\otimes\mathbb{I}_{n})\rho$ 
is defined concerning the canonical product basis as 
\begin{equation}
\bra{ij} (\mathrm{\tau}_{m}\otimes\mathbb{I}_{n})\rho \ket{kl}=\bra{kj} \rho \ket{il}. \nonumber 
\end{equation}
If
\begin{equation}
 (\mathrm{\tau}_{m}\otimes\mathbb{I}_{n})\rho \geq 0,   
\end{equation}
then $\rho$ is either a separable or bound entangled state. This criterion characterizes only the separable states for $2 \times 2$ and $2 \times 3$ systems \cite{Horodecki}. As the general transpose on $M_{mn}(\mathbb{C})$ does not change the determinant of a matrix and the transpose of $(\mathrm{\tau}_{m}\otimes\mathbb{I}_{n})\rho$ is equal to $(\mathbb{I}_{m}\otimes\mathrm{\tau}_{n})\rho$, we obtain that the criterion is independent of the subsystem that is transposed.

\subsection{Volume}

The vector space $M_{mn}(\mathbb{C})$ together with the
Hilbert-Schmidt inner product 
\begin{equation}
\langle X,Y\rangle_{\text{HS}}=\mathrm{Tr} \{X^\dagger Y\} \quad \text{with} \quad X,Y \in M_{mn}(\mathbb{C})    
\end{equation}
becomes a Hilbert space. An elementary orthonormal basis in $M_{mn}(\mathbb{C})$ regarding this inner product is formed by the $mn \times mn$ matrices having only one entry equal to one. The Hilbert-Schmidt norm is induced by the inner product and reads
\begin{equation}
   ||X||^2_{\text{HS}}=\langle X, X \rangle _{\text{HS}}. 
\end{equation}
The distance between two hermitian matrices $X$ and $Y$ is given by
\begin{equation}
  ||X-Y||_{\text{HS}}=\sqrt{\sum^{mn}_{i,j=1} |X_{ij}-Y_{ij}|^2}.  
\end{equation}
The set of self-adjoint matrices forms a subspace in $M_{mn}(\mathbb{C})$ and it is isomorphic to $\mathbb{R}^{mn} \oplus \mathbb{C}^{mn(mn-1)/2}$ or $\mathbb{R}^{(mn)^2}$. Since $\mathbb{C}$ can be identified with $\mathbb{R}^2$, we are dealing with Euclidean vector spaces, where we integrate with respect to the usual Lebesgue measure.

Density matrices form a convex set inside $\mathbb{R}^{(mn)^2-1}$, where the reduction in the dimensionality of the space arises from the constraint $\mathrm{Tr} \{\rho\}=1$. The X-states form inside this convex body another convex body with a lower dimension depending on $m$ and $n$: $mn-1+ 2\lfloor mn/2 \rfloor$, where $\lfloor \cdot \rfloor$ is the floor function. For example, if $m=n=2$ (qubit-qubit)  we have a convex body in $\mathbb{R}^{7}$, and if $m=2$ and $n=3$ (qubit-qutrit) then we deal with $\mathbb{R}^{11}$. Given the nature of X-states and denoting $\mu$ the Lebesque measure on $\mathbb{R}^{mn-1+ 2\lfloor mn/2 \rfloor}$, integrals are computed as 
\begin{equation}
    \int d\mu=\int \prod^{mn-1}_{i=1} d\rho_{ii}  \prod^{mn}_{\substack{i,j=1 \\ i<j \\ i+j=mn+1}} d\operatorname{Re}(\rho_{ij})\, d\operatorname{Im}(\rho_{ij}).
\end{equation}

\subsection{Useful integrals}
In this section, we present the mathematical tools that are essential to the derivation in the following sections.
\subsubsection{Beta function}\label{BetaFunction}
 The beta function is defined as \cite{Davies}
\begin{equation}
    B(a,b) = \int_0^1 x^{a-1} (1-x)^{b-1}\,dx, \quad \operatorname{Re}(a)>0 \quad \operatorname{Re}(b)>0. 
\end{equation} 
In terms of the factorial function, we have
\begin{equation}
    B(a,b)= \frac{\Gamma(a)\Gamma(b)}{\Gamma(a+b)} = \frac{(a-1)!(b-1)!}{(a+b-1)!},
\end{equation}
where $\Gamma(\cdot)$ is the gamma function.

\subsubsection{Dirichlet integral}\label{Dirichlet_CDF}

Let $\Delta=\left\{\left(x_1, \dots, x_d\right): 0 \leq x_i, \, \sum^d_{i=1} x_i \leq 1 \right\}$ be a simplex. We are interested in the Dirichlet integral
\begin{equation}
\label{eq:Dirichlet}
I(\pmb{\alpha}) = \int_{\Delta} x_1^{\alpha_1-1} x_2^{\alpha_2-1} \dots x_d^{\alpha_d-1}\left(1-x_1-\ldots-x_d\right)^{\alpha_0-1} \prod^d_{i=1} dx_i,
\end{equation}
where $\alpha_i>0$ and $\pmb{\alpha}=(\alpha_0, \alpha_1, \dots, \alpha_d)$. The integral has a known expression because it is employed in the definition of the Dirichlet probability density function \cite{Kotz}:
\begin{equation}
I(\pmb{\alpha}) = \frac{\prod_{i=0}^d \Gamma\left(\alpha_i\right)}{\Gamma\left(\sum_{i=0}^d \alpha_i\right)}
\end{equation}

\subsubsection{Dirichlet integral involving a minimum function}

Specifically, we are interested in an integral similar to Eq. \eqref{eq:Dirichlet} but involving a minimum function:
\begin{equation}
    I_0 = \int_0^1 dx_1 \int_0^{1-x_1}dx_2\int_0^{1-x_1-x_2}dx_3 \, \min[x_1(1-x_1-x_2-x_3),x_2x_3]^2.
    \label{eq:Dirichletmin1}
\end{equation}
Processing the minimum function, the integration bound of $x_3$ is divided into two ranges, one in which the minimum is $x_1(1-x_1-x_2-x_3)$ and the other for the minimum being $x_2x_3$. The inequality
\begin{eqnarray}
    x_2x_3 \leq x_1(1-x_1-x_2-x_3)
\end{eqnarray}
leads to
\begin{eqnarray}
&&\int_0^{1-x_1-x_2}dx_3 \, \min[x_1(1-x_1-x_2-x_3),x_2x_3]^2 \\
 &&= \int_0^{\frac{x_1(1-x_1-x_2)}{x_1+x_2}} x_2^2\,x_3^2\,dx_3 + \int_{\frac{x_1(1-x_1-x_2)}{x_1+x_2}}^{1-x_1-x_2}x_1^2\,(1-x_1-x_2-x_3)^2 \,dx_3 \nonumber\\
    &&= \frac{1}{3} \frac{x_1^2x_2^2(1-x_1-x_2)^3}{(x_1+x_2)^2}.\nonumber
\end{eqnarray}
A straightforward calculation yields
\begin{equation}
    I_0 = \frac{1}{12600} = \frac{2}{5}\frac{1}{7!}.
\label{I_0}
\end{equation}

\subsubsection{Incomplete Dirichlet integral with a minimum function}

The final useful integral that is necessary for the later computation is defined as
\begin{eqnarray}
I(n,a) &=& \int_0^a  dx_1\int_0^{a-x_1}  dx_2\int_0^{a-x_1-x_2}  dx_3\int_0^{a-x_1-x_2-x_3} dx_4  \nonumber \\
&\times& \min(x_1x_2,x_3x_4)^2(a - x_1-x_2-x_3-x_4)^n \quad \text{for} \quad a\in[0,1].
\label{eq:Dirichletmin2}
\end{eqnarray}
In fact, this is an integral over the simplex $\left\{\left(x_1, x_2, x_3, x_4\right): 0 \leq x_i, \, \sum^4_{i=1} x_i \leq a  \right\}$ and after change of variables it becomes
\begin{equation}
    I(n,a) = \int_0^a dt \int_0^1 d\alpha \int_0^{1-\alpha}d\beta \int_0^{1-\alpha-\beta}d\gamma \, t^7 \min\left[\alpha\beta, \gamma(1-\alpha-\beta-\gamma)\right]^2 (a-t)^n,
\end{equation}
where $t = x_1 + x_2 + x_3 + x_4$, ranging from $0$ to $a$, as appearing leftmost of the integral and $\alpha = x_1/t, \beta = x_2/t,\gamma = x_3/t$, and finally $x_4 = t-x_1-x_2-x_3 = t(1-\alpha - \beta - \gamma)$. Then, using Eq. \eqref{eq:Dirichletmin1} we obtain
\begin{eqnarray}
     I(n,a) &=& \int_0^a dt\, t^7 (a-t)^n I_0 
     =a^{8+n} I_0  \int_0^1  ds\, s^7 (1-s)^n  \nonumber \\
     &=& a^{8+n} I_0 B(8,n+1).
\end{eqnarray}

\section{Volume of $m \times n$ X-states}
\label{sec:Volumetotal}

To determine the Hilbert-Schmidt inner product-induced volume of the $m \times n$ X-states, we have to evaluate the following integral
\begin{equation}
 V_{m \times n}^X = \int_\Delta \prod^{mn-1}_{i=1} d\rho_{ii}  
 \int \prod^{mn}_{\substack{i,j=1 \\ i<j \\ i+j=mn+1}} d\operatorname{Re}(\rho_{ij})\, d\operatorname{Im}(\rho_{ij}),
    \label{integral_complete}   
\end{equation}
where $\Delta=\left\{\left(\rho_{11},\rho_{22}, \dots \right): 0 \leq \rho_{ii}, \, \sum^{mn-1}_{i=1} \rho_{ii} \leq 1  \right\}$. The second integral is carried over the set of $\operatorname{Re}(\rho_{ij})$ and $d\operatorname{Im}(\rho_{ij})$ defined by the constraint in Eq. \eqref{eq:posconstraint}:
\begin{equation}
\rho_{ii}\rho_{jj} \geq \operatorname{Re}(\rho_{ij})^2 + \operatorname{Im}(\rho_{ij})^2 \quad \text{for} \quad i+j=mn+1,   
\end{equation}
which guarantees the positive semidefinite property of the X-state. Given this inequality, all the integral pairs of $\int d \text{Re}(\rho_{ij}) \int d \text{Im}(\rho_{ij})$ just yield the areas of circles of radius $\sqrt{\rho_{ii}\rho_{jj}}$.

In the case of $mn$ being even, this leads to the expression:
\begin{equation}
    V_{m \times n}^X = \pi^{\frac{mn}{2}} \int_\Delta \prod^{mn-1}_{i=1} d\rho_{ii} \, \rho_{11}\, \rho_{22} \dots \rho_{(mn-1)(mn-1)}\,
    \left(1-\sum_{j = 1}^{mn-1}\rho_{jj}\right) 
\end{equation}
This is a Dirichlet integral discussed in Section \ref{Dirichlet_CDF} with 
\begin{equation}
 \alpha_0=\alpha_1=\dots=\alpha_{mn-1}=2. \nonumber
\end{equation}
Thus,
\begin{equation}
    V_{m \times n}^X =\pi^{\frac{mn}{2}} \frac{1}{(2mn-1)!}, \quad mn = 2k, \quad  k \in \mathbb{N}, \quad k \geq 2.
    \label{even_mn}
\end{equation}

In the case of $mn$ being odd, the middle diagonal element is not involved in any eigenvalue inequalities, which means that it does not appear in the integrand, albeit it as a variable is still being integrated. This means that in terms of the Dirichlet integral, it has an $\alpha$ coefficient of only equal to one, instead of two, yielding the following result:
\begin{equation}
    V_{m \times n}^X = \pi^{\frac{mn-1}{2}} \frac{1}{(2mn-2)!}, \quad mn = 2k+1, \quad k \in \mathbb{N}, \quad k \geq 4.
    \label{odd_mn}
\end{equation}

\section{Volume of $m \times n$ PPT X-states}
\label{sec:VolumePPT}

Upon partial transpose, which is symmetric regarding the two systems concerned, the submatrices segmented either by the first or the second system's levels are transposed. This means that if the transformed matrix was to remain a proper density matrix, its eigenvalues must also remain larger than or equal to zero. For the special case of X-states, the partial transpose exchanges some off-diagonal entries, and we denote their new row and column indices by $i'$ and $j'$. The partially transposed matrix becomes a density matrix, iff
\begin{equation}
\rho_{i'i'}\rho_{j'j'} \geq \operatorname{Re}(\rho_{ij})^2 + \operatorname{Im}(\rho_{ij})^2 \quad \text{for} \quad i'+j'=mn+1,    
\end{equation}
which together with the original constraint
\begin{equation}
 \rho_{ii}\rho_{jj} \geq \operatorname{Re}(\rho_{ij})^2 + \operatorname{Im}(\rho_{ij})^2 \quad \text{for} \quad i+j=mn+1   
\end{equation}
yield
\begin{equation}
    \int d\text{Re}(\rho_{ij}) \int d\text{Im}(\rho_{ij}) = \pi^2 \min(\rho_{ii}\rho_{jj},\rho_{i'i'}\rho_{j'j'})^2
    \label{PPT_int}
\end{equation}
The PPT criterion is a necessary and sufficient condition for the set of separable quantum states in $2 \times 2$ and $2 \times 3$ systems. Therefore, we calculate this system concretely, before the general result is presented. 

\subsection{$2\times 2$ X-states}
\label{sec:2x2}
In the case of a qubit-qubit system, the volume of the X-states is given by Eq. \ref{even_mn}:
\begin{equation}
V_{2\times2}^X = \frac{\pi^2}{7!}.    
\end{equation}
To calculate the volume of separable X-states, one arrives at the following integral with the aid of Eq.\ref{PPT_int}:
\begin{equation}
   V_{2\times2}^{X,PPT} = \int_0^1 d\rho_{11} \int_0^{1-\rho_{11}}d\rho_{22}\int_0^{1-\rho_{11}-\rho_{22}}d\rho_{33} \,\pi^2 \min\left[\rho_{11}(1-\rho_{11}-\rho_{22}-\rho_{33}),\rho_{22}\rho_{33}\right]^2.
\end{equation}
Utilizing the result from Eq. \ref{eq:Dirichletmin1}, we have
\begin{equation}
 V_{2\times2}^{X,PPT}  = \pi^2 I_0 = \pi^2\frac{2}{5}\frac{1}{7!} = \frac{2}{5}V_{2\times2}^X    
\end{equation}
Thus, the ratio between the two volumes for the qubit-qubit system is $\frac{2}{5}$ \cite{Milz}.

\subsection{$2 \times 3$ X-states}
\label{sec:2x3}
The total volume of qubit-qutrit X-states is given by Eq. \ref{even_mn}:
\begin{equation}
 V_{2\times3}^X =  \frac{\pi^2}{11!}.   
\end{equation}
In the case of separable X-states, we have
\begin{eqnarray}
&&V_{2\times3}^{X,PPT}= \int_0^1 d \rho_{22} \int_0^{1-\rho_{22}} d \rho_{55}\,\rho_{22} \rho_{55} \int_0^{1-\rho_{22} - \rho_{55}} d \rho_{11} \int_0^{1-\rho_{11}-\rho_{22} - \rho_{55}} d \rho_{33} \nonumber \\
&& \times \int_0^{1-\rho_{11}-\rho_{22} -\rho_{33}- \rho_{55}} d \rho_{44} \, \pi^2 \min\left[\rho_{33} \rho_{44}, \rho_{11}\left(1-\sum_{i=1}^5 \rho_{ii}\right)\right]^2 
\nonumber \\
&&= \pi^2 \int_0^1 d \rho_{22} \int_0^{1-\rho_{22}} d \rho_{55}\,\rho_{22} \rho_{55} (1-\rho_{22}-\rho_{55})^7 I_0 \nonumber \\
&&=\pi^2 I[\pmb{\alpha}=(2,2,8)]I_0 =\pi^2 \frac{\Gamma(2)\Gamma(2) \Gamma(8)}{\Gamma(12)} \frac{2}{5} \frac{1}{7!}=\frac{2}{5} V_{2 \times 3}^X.
\end{eqnarray}
In the above derivation, we have employed the Dirichlet integrals of both Eqs.~\eqref{eq:Dirichlet} and~\eqref{eq:Dirichletmin1}. Furthermore, the integration over the simplex $\Delta=\left\{\left(\rho_{11}, \dots, \rho_{55} \right): 0 \leq \rho_{ii}, \, \sum^{5}_{i=1} \rho_{ii} \leq 1  \right\}$ was represented in such a way that first those diagonal elements are integrated, which appeared in the minimum function. The conclusion is that, similarly to the qubit-qubit scenario, the volume ratio remains $\frac{2}{5}$.

\subsection{$m \times n$ PPT X-states}
Let $A$ be the number of minimum functions concerning four diagonal elements, $B$ be the number of pairs of diagonal elements entering the final integral linearly, and finally, $C$, which is either zero if $mn$ is even or one if $mn$ is odd, indicating that a non-participating diagonal element that is right at the cross point of the X-state density matrix. The expressions of $A$, $B$ and $C$ in terms of $m$ and $n$ are:

\begin{center}
\begin{tabular}{|c|c|c|}
\hline
     & odd $m$ & even $m$  \\
     \hline
   odd $n$ & $A = \frac{(m-1)(n-1)}{4}$, $B = \frac{m+n-2}{2}$, $C = 1$ &  $A = \frac{m(n-1)}{4}$, $B = \frac{m}{2}$, $C = 0$\\
   \hline
   even $n$ & $A = \frac{(m-1)n}{4}$, $B = \frac{n}{2}$, $C = 0$ & $A = \frac{mn}{4}$, $B = C = 0$ \\
   \hline
\end{tabular}
\end{center}
They fulfill the condition 
\begin{equation}
4A + 2B + C = mn.    
\end{equation}

We denote $\mathbb{A}$ the set of $4$-tuples $(i,j,k,l)$ with $i+j=k+l=mn+1$ and $1\leq i,j,k,l \leq mn$, for which the off-diagonal entry $\rho_{ij}$ is swapped with $\rho_{kl}$ during the partial transpose. In this manner, $\mathbb{B}$ is the set of $2$-tuples $(i,j)$ with $i+j=mn+1$ and $1\leq i,j \leq mn$, for which the off-diagonal entry $\rho_{ij}$ is not affected by the partial transpose. Then, the volume of $m \times n$ PPT X-states for $mn$ being even reads
\begin{equation}
V_{m\times n}^{X,PPT}=\pi^{\frac{mn}{2}} \int_{\Delta}  \prod_{(i,j)\in \mathbb{B}} (\rho_{ii} \rho_{jj}) 
\prod_{(i,j,k,l) \in \mathbb{A} } \min(\rho_{ii} \rho_{jj},\rho_{ll} \rho_{kk})^2 \,\prod_{i=1}^{mn-1}d\rho_{ii}.
\label{eq:genPPT}
\end{equation}

As we have seen in the $2 \times 3$ case, keeping the lower indices of the rows and columns may complicate the integration over the simplex $\Delta=\left\{\left(\rho_{11}, \rho_{22}, \dots, \right): 0 \leq \rho_{ii}, \, \sum^{nm-1}_{i=1} \rho_{ii} \leq 1  \right\}$.  However, the simplex is invariant under the permutation of its coordinates, which can be seen as renaming them. Therefore, it is worth renaming the lower indices in such a way that those diagonal entries $\rho_{ii}$, which are taking part in a minimum function are at the end when $i$ takes values from $1$ to $mn-1$. In this sense, if $\mathbb{B}$ is not an empty set, contains the $2$-tuples like $(1,2)$. Let us denote the largest index in $\mathbb{B}$ by $b$, which means that there is an element $(b-1,b)$ in the set. Now, if $\mathbb{B} \neq \emptyset$, Eq. \eqref{eq:genPPT} can be rewritten as
\begin{eqnarray}
\label{eq:mneven1}
&&V_{m\times n}^{X,PPT}=\pi^{\frac{mn}{2}} \int^1_0 
d\rho_{11} \dots  \int^{1-\sum^{mn-1}_{i=1}\rho_{ii}}_0 d\rho_{(mn-1)(mn-1)} \prod^{b}_{i=1} \rho_{ii}  \\
&&\times  \min\left[\rho_{(b+1)(b+1)} \rho_{(b+2)(b+2)},\rho_{(b+3)(b+3)} \rho_{(b+4)(b+4)}\right]^2
\dots \nonumber \\
&&\times  \min\left[\rho_{(mn-3)(mn-3)} \rho_{(mn-2)(mn-2)},\rho_{(mn-1)(mn-1)} \left(1-\sum^{mn-1}_{i=1}\rho_{ii}\right)\right]^2, \nonumber
\end{eqnarray}
otherwise as 
\begin{eqnarray}
\label{eq:mneven2}
&&V_{m\times n}^{X,PPT}= \\
&&=\pi^{\frac{mn}{2}} \int^1_0 
d\rho_{11} \dots  \int^{1-\sum^{mn-1}_{i=1}\rho_{ii}}_0 d\rho_{(mn-1)(mn-1)} \,  \min\left[\rho_{11} \rho_{22},\rho_{33} \rho_{44}\right]^2 \dots \nonumber \\
&&\times  \min\left[\rho_{(mn-3)(mn-3)} \rho_{(mn-2)(mn-2)},\rho_{(mn-1)(mn-1)} \left(1-\sum^{mn-1}_{i=1}\rho_{ii}\right)\right]^2. \nonumber
\end{eqnarray}
The integrals of the minimum functions involve $4A-1$ variables. The integral with respect to $\rho_{(mn-3)(mn-3)}$, $\rho_{(mn-2)(mn-2)}$, and $\rho_{(mn-1)(mn-1)}$ reads
\begin{eqnarray}
   && \int^{1-\sum^{mn-4}_{i=1}\rho_{ii}}_0 d\rho_{(mn-3)(mn-3)} \dots \int^{1-\sum^{mn-2}_{i=1}\rho_{ii}}_0 d\rho_{(mn-1)(mn-1)} \nonumber \\
   && \times \min\left[\rho_{(mn-3)(mn-3)} \rho_{(mn-2)(mn-2)},\rho_{(mn-1)(mn-1)} \left(1-\sum^{mn-1}_{i=1}\rho_{ii}\right)\right]^2 \nonumber \\
   &&=\left(1-\sum^{mn-4}_{i=1}\rho_{ii} \right)^7 I_0,
\end{eqnarray}
where after a change of variables similar to the one employed in Sec. \ref{sec:2x3} we have used the Eq. \eqref{eq:Dirichletmin1}. For the remaining $4A-4$ variables, we recognize first that
\begin{eqnarray}
  && \int^{1-\sum^{mn-8}_{i=1}\rho_{ii}}_0 d\rho_{(mn-7)(mn-7)} \dots \int^{1-\sum^{mn-5}_{i=1}\rho_{ii}}_0 d\rho_{(mn-4)(mn-4)} \nonumber \\
   && \times \min\left[\rho_{(mn-7)(mn-7)} \rho_{(mn-6)(mn-6)},\rho_{(mn-5)(mn-5)} \right]^2 \left(1-\sum^{mn-4}_{i=1}\rho_{ii} \right)^7 I_0 \nonumber \\   
   &&=I\left(7, 1- \sum^{mn-8}_{i=1}\rho_{ii}\right) I_0=
   \left(1- \sum^{mn-8}_{i=1}\rho_{ii}\right)^{15} B(8,8)I^2_0,
\end{eqnarray}
where we have used Eq. \eqref{eq:Dirichletmin2}. The next integral with respect to $\rho_{(mn-11)(mn-11)}$, $\rho_{(mn-10)(mn-10)}$, $\rho_{(mn-9)(mn-9)}$, and $\rho_{(mn-8)(mn-8)}$ has a similar structure as Eq. \eqref{eq:Dirichletmin2}. Thus, Eq. \eqref{eq:mneven1} reads
\begin{eqnarray}
&&V_{m\times n}^{X,PPT}=\pi^{\frac{mn}{2}} \int^1_0 
d\rho_{11} \dots  \int^{1-\sum^{b-1}_{i=1}\rho_{ii}}_0 d\rho_{bb} \prod^{b}_{i=1} \rho_{ii}  \\
&&\times \left(1-\sum^{b}_{i=1} \rho_{ii}\right)^{8(A-1)+7} B(8,8A-8) \dots B(8,8) I^A_0 \nonumber \\
&&=\pi^{\frac{mn}{2}}\frac{\Gamma(8A)}{\Gamma(8A+4B)} \frac{\Gamma(8) \Gamma(8A-8)}{\Gamma(8A)}\dots  \frac{\Gamma(8) \Gamma(8)}{\Gamma(16)} \left(\frac{2}{5 \Gamma(8)}\right)^A \nonumber \\
&&=\left(\frac{2}{5}\right)^A \pi^{\frac{mn}{2}} \frac{1}{(2mn-1)!} = \left(\frac{2}{5}\right)^A V_{m \times n}^X,
\end{eqnarray}
where we have used the integral in Eq. \eqref{eq:Dirichlet} with $\pmb{\alpha}=(2, 2, \dots, 8A)$
and $8A+4B=2mn$, because $mn$ is even. Similarly, for Eq. \eqref{eq:mneven2} we have
\begin{eqnarray}
&&V_{m\times n}^{X,PPT}=\pi^{\frac{mn}{2}} B(8,8A-8) \dots B(8,8) I^A_0 \nonumber \\
&&=\pi^{\frac{mn}{2}} \frac{\Gamma(8) \Gamma(8A-8)}{\Gamma(8A)}\dots  \frac{\Gamma(8) \Gamma(8)}{\Gamma(16)} \left(\frac{2}{5 \Gamma(8)}\right)^A \nonumber \\
&&=\left(\frac{2}{5}\right)^A \pi^{\frac{mn}{2}} \frac{1}{(2mn-1)!} = \left(\frac{2}{5}\right)^A V_{m \times n}^X,
\end{eqnarray}
where $8A=mn$, because $B=0$.

In the case, when $C=1$, Eq. \eqref{eq:genPPT} can be rewritten as
\begin{eqnarray}
&&V_{m\times n}^{X,PPT}=\pi^{\frac{mn-1}{2}} \int^1_0 
d\rho_{11} \dots  \int^{1-\sum^{mn-1}_{i=1}\rho_{ii}}_0 d\rho_{(mn-1)(mn-1)} \prod^{b}_{i=2} \rho_{ii}  \\
&&\times  \min\left[\rho_{(b+1)(b+1)} \rho_{(b+2)(b+2)},\rho_{(b+3)(b+3)} \rho_{(b+4)(b+4)}\right]^2
\dots \nonumber \\
&&\times  \min\left[\rho_{(mn-3)(mn-3)} \rho_{(mn-2)(mn-2)},\rho_{(mn-1)(mn-1)} \left(1-\sum^{mn-1}_{i=1}\rho_{ii}\right)\right]^2, \nonumber \\
&&=\pi^{\frac{mn-1}{2}}\frac{\Gamma(8A)}{\Gamma(8A+4B+1)} \frac{\Gamma(8) \Gamma(8A-8)}{\Gamma(8A)}\dots  \frac{\Gamma(8) \Gamma(8)}{\Gamma(16)} \left(\frac{2}{5 \Gamma(8)}\right)^A \nonumber \\
&&=\left(\frac{2}{5}\right)^A \pi^{\frac{mn-1}{2}} \frac{1}{(2mn-2)!} = \left(\frac{2}{5}\right)^A V_{m \times n}^X,
\end{eqnarray}
because $8A+4B+2=2mn$.

Finally, we rewrite $A$ to obtain
\begin{equation}
    \frac{V_{m\times n}^{X,PPT}}{V_{m \times n}^X}=\left(\frac{2}{5}\right)^A=0.4^{\lfloor \frac{m}{2} \rfloor \cdot \lfloor \frac{n}{2} \rfloor}.
\end{equation}
Apart from the $2 \times 2$ and $2 \times 3$ cases, this ratio is an upper bound for the volume ratio between separable and all X-states.

\section{Concluding remarks}
\label{sec:conluding}

In this paper, we have proved that the volume of the PPT X-states in $m \times n$ systems decreases exponentially as a function of the dimensions. This behavior was expected based on the result of Ref. \cite{infinite} on infinite dimensional Hilbert spaces, however, the exponent involves the floor functions of $m/2$ and $n/2$. This indicates that further future extensions of this research have to take into account the parities of the dimensions. It is important to point out that the result holds for the Hilbert-Schmidt measure.  

To compute integrals over the state space, we have employed Dirichlet-type integrals and exploited the symmetric structure of the eigenvalues of the X-states, which led us to a set of constraints on the matrix entries.
The partial transpose does not change the visual appearance of an X-state, its trace, and hermitianity. Therefore, only new constraints to secure the positive semidefinite property have been obtained. Finally, we have investigated the three possible cases concerning the parities of $m$ and $n$.

\section{Acknowledgement}
This work was supported by AIDAS-AI, Data Analytics and Scalable Simulation, which is a Joint Virtual Laboratory gathering the Forschungszentrum Jülich and the
French Alternative Energies and Atomic Energy Commission. Y.X.W. is grateful to Mees Hendriks for stimulating discussions and thanks the domestic animals dedicated to mental support for scientific research. 

\end{document}